\documentclass[aps,prb,reprint,longbibliography,citeautoscript,floatfix]{revtex4-2}
\usepackage[utf8]{inputenc}
\usepackage[T1]{fontenc}

\usepackage{graphicx} %
\usepackage{siunitx}
\usepackage{float}
\usepackage[normalem]{ulem}
\usepackage{dcolumn} %
\usepackage{microtype}

\newfloat{suppfig}{H}{losuppfig}
\floatname{suppfig}{\textsc{Supplemental Fig.}}

\newfloat{supptab}{H}{losupptab}
\floatname{supptab}{\textsc{Supplemental Table}}

\newcommand{\mytitle}{Stable nanofacets in [111] tilt grain boundaries of face-centered cubic metals}

\usepackage{hyperref} %
\hypersetup{
  final,
  pdfborder={0 0 0},
  pdftitle={\mytitle},
  pdfauthor={Brink, Langenohl, Pemma, Liebscher, Dehm},
  colorlinks=true,
  urlcolor=blue,
  linkcolor=blue,
  citecolor=blue
}

\begin{document}
\frenchspacing

\title{\mytitle}

\author{Tobias Brink}
\email{t.brink@mpie.de}
\thanks{T.B. and L.L. contributed equally.}
\affiliation{Max-Planck-Institut f\"ur Eisenforschung GmbH,
  Max-Planck-Stra\ss{}e 1, 40237 D\"usseldorf, Germany}

\author{Lena Langenohl}
\thanks{T.B. and L.L. contributed equally.}
\affiliation{Max-Planck-Institut f\"ur Eisenforschung GmbH,
  Max-Planck-Stra\ss{}e 1, 40237 D\"usseldorf, Germany}

\author{Swetha Pemma}
\affiliation{Max-Planck-Institut f\"ur Eisenforschung GmbH,
  Max-Planck-Stra\ss{}e 1, 40237 D\"usseldorf, Germany}

\author{Christian H. Liebscher}
\affiliation{Max-Planck-Institut f\"ur Eisenforschung GmbH,
  Max-Planck-Stra\ss{}e 1, 40237 D\"usseldorf, Germany}

\author{Gerhard Dehm}
\affiliation{Max-Planck-Institut f\"ur Eisenforschung GmbH,
  Max-Planck-Stra\ss{}e 1, 40237 D\"usseldorf, Germany}

\date{\today}

\begin{abstract}
  Grain boundaries can dissociate into facets if that reduces their
  excess energy. This, however, introduces line defects at the facet
  junctions, which present a driving force to grow the facets in order
  to reduce the total number of junctions and thus the system's
  energy. Often, micrometer-sized facet lengths are observed and facet
  growth only arrests for kinetic reasons. So far, energetically
  stable, finite-sized facets have not been observed, even though
  theoretical stability conditions have already been proposed. Here,
  we show a case where nanometer-sized facets are indeed stable
  compared to longer facets in $[11\overline{1}]$ tilt grain
  boundaries in Cu by atomistic simulation and transmission electron
  microscopy.  The facet junctions lack a Burgers vector component,
  which is unusual, but which removes the main energy cost of facet
  junctions. Only attractive interactions via line forces remain,
  which result from a discontinuity of grain boundary excess stress at
  the junction. Atomistic simulations predict that the same phenomenon
  also occurs in at least Al and Ag.
\end{abstract}

\maketitle

\newcounter{supplfigctr}
\renewcommand{\thesupplfigctr}{S\arabic{supplfigctr}}
\newcounter{suppltabctr}
\renewcommand{\thesuppltabctr}{S-\Roman{suppltabctr}}
{\refstepcounter{supplfigctr}\label{fig:suppl:S19b:domino}}
{\refstepcounter{supplfigctr}\label{fig:suppl:S19b:zipper}}
{\refstepcounter{supplfigctr}\label{fig:suppl:S19b:overlay-zipper-over-domino}}
{\refstepcounter{supplfigctr}\label{fig:suppl:S37c:domino}}
{\refstepcounter{supplfigctr}\label{fig:suppl:S37c:zipper}}
{\refstepcounter{supplfigctr}\label{fig:suppl:S37c:overlay-zipper-over-domino}}
{\refstepcounter{supplfigctr}\label{fig:suppl:S61b:domino}}
{\refstepcounter{supplfigctr}\label{fig:suppl:S61b:zipper}}
{\refstepcounter{supplfigctr}\label{fig:suppl:S61b:overlay-zipper-over-domino}}
{\refstepcounter{supplfigctr}\label{fig:suppl:S127:domino}}
{\refstepcounter{supplfigctr}\label{fig:suppl:S127:zipper}}
{\refstepcounter{supplfigctr}\label{fig:suppl:S127:overlay-zipper-over-domino}}
{\refstepcounter{supplfigctr}\label{fig:suppl:S169b:domino}}
{\refstepcounter{supplfigctr}\label{fig:suppl:S169b:zipper}}
{\refstepcounter{supplfigctr}\label{fig:suppl:S169b:overlay-zipper-over-domino}}
{\refstepcounter{supplfigctr}\label{fig:suppl:S3:zipper}}
{\refstepcounter{supplfigctr}\label{fig:suppl:facet-energies-symm-asymm}}
{\refstepcounter{supplfigctr}\label{fig:suppl:Burgers-junction:others}}
{\refstepcounter{supplfigctr}\label{fig:suppl:S169-lattice-strains}}
{\refstepcounter{suppltabctr}\label{tab:suppl:excess-properties}}
{\refstepcounter{supplfigctr}\label{fig:suppl:excess-properties}}
{\refstepcounter{suppltabctr}\label{tab:suppl:excess-properties-per-plane}}
{\refstepcounter{supplfigctr}\label{fig:suppl:excess-properties-per-plane}}
{\refstepcounter{supplfigctr}\label{fig:suppl:trapezoidal-stress}}
{\refstepcounter{supplfigctr}\label{fig:suppl:zipper-Burgers-circuits}}
{\refstepcounter{supplfigctr}\label{fig:suppl:zipper-energy}}

\section{Introduction}

Grain boundaries (GBs) are planar defects in crystalline solids and
thus lead to an excess (free) energy. Oftentimes, GBs nevertheless
remain in the system for kinetic reasons, but tend to reduce their
excess energy locally. For example, if the GB energy is sufficiently
anisotropic, GBs are known to decompose into energetically favorable
facets. Typically, asymmetric GBs split into symmetric facets
\cite{Wagner1974, Brokman1981}, but transitions from one symmetric GB
plane into facets of different symmetric GB planes are also known,
most famously for $\Sigma3$ $[11\overline{1}]$ tilt GBs with $\{011\}$
habit planes in fcc metals \cite{Hsieh1989, Straumal2001,
  Hamilton2003, Wu2009, Banadaki2016}. Faceting/defaceting transitions
as a function of temperature were reported \cite{Hsieh1989, Wu2009,
  Straumal2016}. However, once facets appear, there is a driving force
for their growth \cite{Hamilton2003, Wu2009, Hadian2018}. This is
because the line separating different facets---the facet junction---is
a defect \cite{Dimitrakopulos1997, Pond1997, Medlin2017} and the
system can reduce its energy by reducing the number of junctions
\cite{Hamilton2003, Wu2009}. Nanoscale facet sizes have been observed
\cite{Medlin2017, Peter2018}, but there the facet junctions are most
likely kinetically pinned by immobile GB defects \cite{Medlin2017} or
segregated elements \cite{Peter2018}.

\begin{figure}
    \centering
    \includegraphics{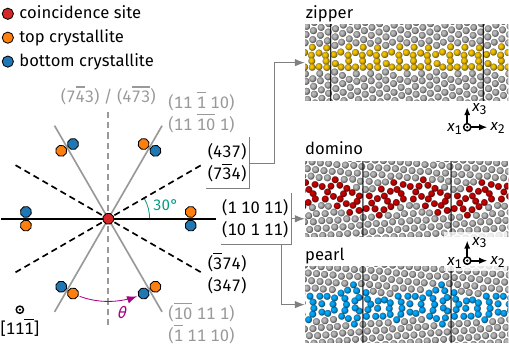}
    \caption{Crystallography of $\Sigma$37c $[11\overline{1}]$ tilt
      GBs and their GB phases. An excerpt of the dichromatic pattern
      around a coincidence site shows the possible symmetric GB planes
      (lines): Solid lines exhibit the domino and pearl phases, while
      dashed lines exhibit the zipper phases, as shown in the
      snapshots on the right. Here, the pearl phase will not be
      discussed further, as it is not relevant for nanofaceting. For
      the GB habit planes, Miller indices are listed for the respective
      crystallographic planes of the top and bottom crystallite (in
      that order). Equivalent planes occur every \ang{60} due to the
      three-fold symmetry of the fcc crystal. The two variants of
      symmetric planes are inclined by \ang{30}. For simplicity, we
      concentrate on the GB planes in black and not on the
      crystallographically equivalent gray planes. Our convention for
      the simulation cell coordinate systems is
      $x_1 \parallel [11\overline{1}]$, $x_2$ lying in the GB plane,
      and $x_3$ being the GB plane normal. Axes $x_2$ and $x_3$ are
      different crystallographic directions for each GB (compare,
      e.g., zipper with domino). The other tilt GBs in this paper have
      equivalent crystallography for their respective misorientation
      $\theta$.\vspace{-2\baselineskip}}
    \label{fig:geometry}
\end{figure}

Here, we report on more general $[11\overline{1}]$ tilt GBs in fcc
metals. The $\Sigma19$b $[11\overline{1}]$ $\{178\}$ (misorientation
$\theta = \ang{46.83}$) and $\Sigma37$c $[11\overline{1}]$
$\{1\,10\,11\}$ GBs ($\theta = \ang{50.57}$) in Cu and Al are close to
the $\Sigma3$ $[11\overline{1}]$ $\{011\}$ GB ($\theta = \ang{60}$),
but do not exhibit macroscopic faceting \cite{Meiners2020,
  Langenohl2022, Ahmad2024}. Instead, an ordered GB structure termed
domino was found \cite{Meiners2020, Langenohl2022, Brink2023,
  Ahmad2024}. In Cu, an additional pearl structure occurs
\cite{Meiners2020, Langenohl2022, Brink2023}, meaning two GB phases
\cite{Hart1968, Cahn1982, Rottman1988, Frolov2015a, Cantwell2020}
exist and can transition into one another based on the thermodynamic
conditions \cite{Meiners2020, Langenohl2022, Brink2023}. On the
second symmetric plane, inclined by \ang{30}, only a single GB phase
(termed zipper) was reported for $\Sigma19$b $[11\overline{1}]$
$\{235\}$ and $\Sigma37$c $[11\overline{1}]$ $\{347\}$ GBs
\cite{Meiners2020a, Langenohl2023,
  Ahmad2024}. Figure~\ref{fig:geometry} shows these GB phases and the
crystallography of the $\Sigma37$c GB as an example.

In this paper, we will focus on the relation between the domino and
zipper phases and will not discuss the pearl phase in detail: We
demonstrate that the domino phase does in fact consist of zipper
structure facets and that its facet junctions have an attractive
interaction. This means that the domino phase's GB energy is lower for
shorter facet lengths and that the observed nanofacets are
energetically stable.

\section{Methods}
\subsection{Simulation}

We used \textsc{lammps} \cite{Plimpton1995, Thompson2022} for
molecular statics and molecular dynamics (MD) simulations of
$[11\overline{1}]$ tilt GBs using embedded atom method (EAM)
potentials for Cu \cite{Mishin2001}, Al \cite{Mishin1999}, and Ag
\cite{Williams2006}. Bicrystals were constructed by joining two
crystallites in the desired orientation, sampling different relative
displacements, and minimizing atomic positions with regards to the
energy until the desired GB phases were found ($T =
\SI{0}{K}$). Details are provided in previous publications
\cite{Brink2023, Langenohl2023}. GB planes and $\Sigma$ values as
function of the misorientation were found with the code from
Ref.~\cite{Hadian2018a}. As a convention, we call the tilt axis
direction $x_1$, its normal along the (average) GB plane $x_2$, and
the GB plane normal $x_3$.

We observed the evolution of GB facets using MD simulations with a
time integration step of \SI{2}{fs}. Temperature was controlled in the
whole system with Nos\'e--Hoover thermostats. We used periodic
boundary conditions along the tilt axis $x_1$ and open boundaries
normal to the GB in $x_3$ direction. We also used periodic boundaries
in $x_2$ direction where not specified otherwise. The simulation cell
length in periodic directions was fixed according to the relevant fcc
lattice constant at the target temperature. We visualized MD
simulations with \textsc{ovito} \cite{Stukowski2010} and distinguished
between fcc and GB atoms using the polyhedral template matching method
\cite{Larsen2016}.

Raw (meta-)data for the simulations are provided in the companion
dataset \cite{zenodo}.

\subsection{Experiment}

We additionally conducted experiments on Cu thin films that were
epitaxially grown on $\langle0001\rangle$ Al$_2$O$_3$ wafers. The
films were subsequently annealed for at least \SI{1}{h} at 673 to
\SI{723}{K} \cite{Langenohl2022, Langenohl2023}. TEM lamellas
containing GBs with different misorientation angles and GB planes were
extracted using the focused Ga$^+$ ion beam of a Thermo Fisher
Scientific Scios2HiVac dual-beam secondary electron microscope. They
were then investigated in a FEI Titan Themis 80-300 (Thermo Fisher
Scientific) scanning transmission electron microscope (STEM) at a
voltage of \SI{300}{kV}, beam currents of 70 to \SI{80}{pA}, and a
convergence angle of \SI{17.8}{mrad}. A high-angle annular dark field
(HAADF) STEM detector (Fishione Instruments Model 3000) with a
collection angle of 78 to \SI{200}{mrad} was used for the registration
of all images. The interested reader is referred to
Refs.~\cite{Langenohl2022, Langenohl2023} for more details on the
experimental
methods.%

Raw (meta-)data for the experiments are provided in the companion
dataset \cite{zenodo}.

\subsection{Dislocation-like defects on GBs and construction of Burgers circuits}
\label{sec:methods-Burgers}

To investigate the GB facet junctions---which are line defects---we
first define how to calculate the Burgers vector content of GB
defects. In general, a Burgers vector is a line integral over the
elastic strain field as \cite{HirthLothe1992}
\begin{equation}
  \label{eq:burgers}
  \mathbf{b} = \oint_C \frac{\partial\mathbf{u}}{\partial l} \mathrm{d}l,
\end{equation}
where $C$ is a closed line around the defect of interest, $\mathbf{u}$
is the elastic displacement field, and $l$ is the integration variable
along the line $C$. On a crystal lattice, this can be discretized by
using $\mathbf{u} = \mathbf{x} - \mathbf{X}$, where $\mathbf{x}$ are
the actual atomic coordinates and $\mathbf{X}$ the atomic coordinates
in the defect-free reference system. If we do the integration from one
atom to the next, using steps of $\Delta \mathbf{x}$ (with
corresponding $\Delta \mathbf{X}$), we arrive at \cite{HirthLothe1992}
\begin{equation}
  \label{eq:burgers-discrete}
  \mathbf{b} = \sum_{i\in C} \Delta \mathbf{X}_i.
\end{equation}
For GB defects (e.g., disconnections/secondary GB dislocations), we
simply define the reference coordinates $\mathbf{X}$ in terms of a
system that already contains the GB that we want to treat as defect
free. As a consequence a defect is always defined in relation to a
chosen reference system, which is useful when treating defects of
defects, e.g. line defects of GBs. We here ignore the intrinsic
(geometric) dislocation content of a specific GB and only ask how does
the total dislocation content of a defective GB differ from what we
define as a defect-free GB.

We utilize a simple method in the form used by Medlin et al.\
\cite{Medlin2017}. Here, the crystals abutting the GB are indexed and
a closed loop around the GB defect is formed by going from atom to
atom in steps of $\Delta\mathbf{X}$. In each of the two crystallites
$\lambda$ and $\mu$, the steps $\Delta\mathbf{X}$ are summed up for a
partial circuit that is fully contained in the GB-free parts of the
respective crystallite. The closed line $C$ should then only be
missing two GB crossings, for which we cannot easily access
$\mathbf{X}$. If these crossings are chosen at crystallographically
equivalent sites, however, the sum of these two crossings equals
$\mathbf{0}$ and they can be ignored for obtaining
$\mathbf{b}$. Finally, the two crystal lattices are rotated towards
each other \cite{Pond1989}, here around the tilt axis, as just adding
up the two partial circuits would also contain the primary dislocation
content of the GB (compare Read and Shockley's equation
\cite{Read1950}). Thus, the partial circuits
$\sum_{i\in\lambda} \Delta \mathbf{X}_i$ and
$\sum_{i\in\mu} \Delta \mathbf{X}_i$ are rotated into a common
coordinate system before summing, thus obtaining only the additional
Burgers vector content \cite{Pond1989, Medlin2017}. Usually, we would use a
rotation with the misorientation angle $\theta$ for the given tilt GB
to perform this coordinate transformation. However, we can in
principle also use the misorientation angle of another GB. This then
allows us to locally extract the difference of Burgers vector content
between different misorientations, e.g., to describe a $\Sigma37$c GB
as a $\Sigma3$ GB with defects.

For GB facet junctions, the two GB crossing cannot be made equivalent
because both facets are rotated against each other \cite{Ahmad2024}. For
such cases, Frolov et al.\ proposed a method where $\Delta \mathbf{X}$ of
the GB crossings are measured directly in a reference system that has
to be simulated separately \cite{Frolov2021}. This is then added to
the partial circuits from the previous method.

\section{Results}
\subsection{Stable nanofaceting}

\begin{figure}[t!]
    \centering
    \includegraphics{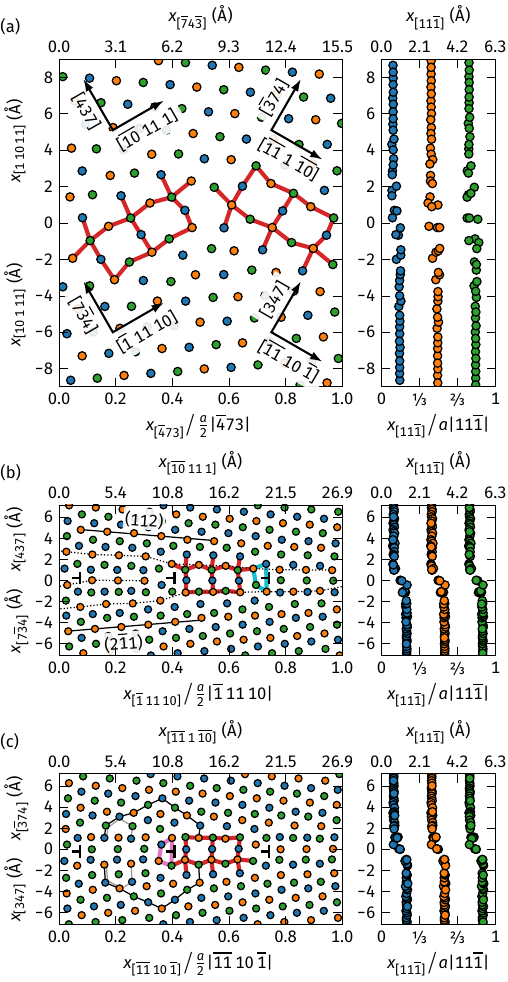}
    \caption{Motifs of $\Sigma37$c GBs in Cu. (a) Domino consists of a
      (b) left and (c) right zipper motif. Atom color indicates ABC
      stacking. Red lines are used to highlight the square motifs,
      blue and pink lines each highlight one trapezoidal unit. Black
      lines in (b) follow $\{112\}$ planes, which is the GB plane of
      the corresponding $\Sigma3$ GB, while black/gray lines in (c)
      indicate Burgers circuits. As discussed in
      Sec.~\ref{sec:methods-Burgers}, the local dislocation content
      with reference to the $\Sigma3$ $[11\overline{1}]$ $\{112\}$ GB
      can be obtained using a modification of the method employed by
      Medlin et al.\ \cite{Medlin2017}. We obtain
      $\mathbf{b} = a/6\langle 112\rangle$ for the black circuit and
      \textbf{0} for the gray one, indicating that the trapezoidal
      motif has dislocation character. Some more technical details are
      provided in Supplemental
      Fig.~\ref*{fig:suppl:zipper-Burgers-circuits}.}
    \label{fig:motifs}
\end{figure}

\begin{figure}
    \centering
    \includegraphics{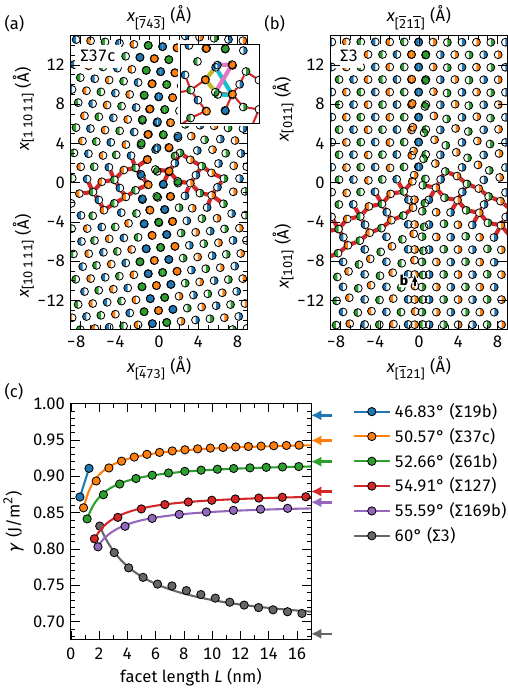}
    \caption{Facets in Cu. (a) Joining two $\Sigma37$c zippers without
      further relaxation. The atoms coming from the left zipper are
      indicated by circles filled on the left half and vice versa for
      the right zipper. In the center region, atoms from both
      structures are shown and overlap. Thus there is no Burgers
      vector. The inset shows how the trapezoidal units indicated in
      Fig.~\ref{fig:motifs} overlap seamlessly. The yellow lines are
      shared between both trapezoidal units. (b) For $\Sigma3$,
      however, the indicated Burgers vector remains, here seen as a
      shift in the bottom crystal. (c) GB energies of GBs with an
      average plane corresponding to domino but with different facet
      lengths $L$. The arrows indicate the GB energy of the zipper
      phase projected onto the domino GB plane (first term of
      Eq.~\ref{eq:faceted-gamma}). This corresponds to infinite facet
      lengths. For $\Sigma19$b, longer facets became instable even in
      static minimization, reverting partially to domino and are not
      shown here.}
    \label{fig:facet-energies}
\end{figure}

The $\Sigma37$c $[11\overline{1}]$ $\{1\,10\,11\}$ domino phase from
atomistic simulations of Cu is shown in Fig.~\ref{fig:motifs}(a). Red
lines highlight an arbitrary choice of atomic motifs---which we simply
call ``squares''---serving to guide the eye. It appears that the two
motifs are each inclined by \ang{\pm 30} towards the GB plane,
coinciding with the indicated crystal directions of the inset
coordinate systems. They correspond to the habit planes of the zipper
structure in the $\Sigma37$c $[11\overline{1}]$ $\{347\}$ GB
[Figs.~\ref{fig:motifs}(b)--(c)]. We can find the same square motifs
from the domino phase in the zipper phase. By rotation of the zipper
structures around $[11\overline{1}]$ by \ang{\pm30}, we can combine
the zipper structures into the domino phase without introducing
long-ranged lattice distortions (Fig.~\ref{fig:facet-energies}(a) and
Supplemental Fig.~\ref*{fig:suppl:S37c:overlay-zipper-over-domino} in
Ref.~\cite{suppl}). The same is found for other misorientation angles
we investigated ($\theta = \ang{46.83}$ to \ang{55.59}, see
Supplemental
Figs.~\ref*{fig:suppl:S19b:domino}--\ref*{fig:suppl:S169b:overlay-zipper-over-domino}). This
suggests that the domino phase can be considered as a nanofaceted
variant of the zipper structure. The structures in Al and Ag are the
same \cite{Brink2023} and this result consequently also applies to
these metals. In previous experiments \cite{Meiners2020,
  Langenohl2022, Ahmad2024} and simulations \cite{Meiners2020,
  Langenohl2022, Brink2023}, however, facet growth was never observed
for domino. This is atypical, since---for example---the related
$\Sigma3$ $[11\overline{1}]$ $\{011\}$ is prone to facet formation and
growth (at least at low temperatures) \cite{Hsieh1989, Straumal2001,
  Hamilton2003, Wu2009, Banadaki2016, Ahmad2024}. Consequently, it is
possible that the domino phase is not faceted, but simply related to
the zipper structure.

\begin{figure}
  \centering
  \includegraphics[]{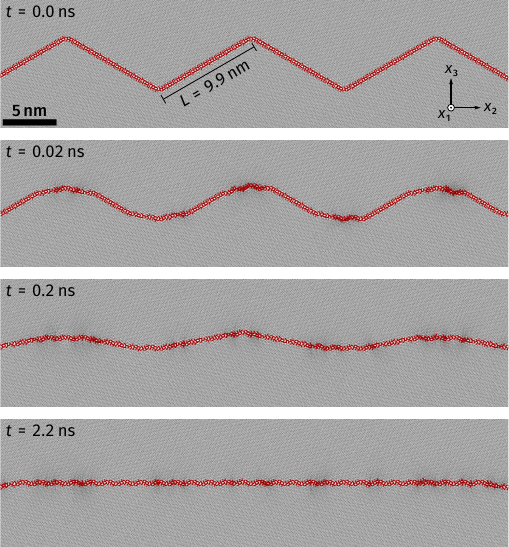}
  \caption{MD simulation of a $\Sigma37$c GB with initial facet length
    of $L = \SI{9.9}{nm}$ at $T = \SI{500}{K}$. Gray atoms were
    identified by polyhedral template matching as fcc atoms, while red
    atoms represent non-fcc atoms, i.e., the GB. The facets almost
    immediately start shrinking at the junction. This also indicates
    that even macroscopic facets would be able to shrink to
    nanofacets, because the process starts on the nanoscale at the
    junctions---where nanofaceting can immediately reduce the GB
    energy---and then proceeds along the GB.}
  \label{fig:md-shrinkage}
\end{figure}
To clarify this, we investigated the GB energies when artificially
constructing longer facets for different misorientation angles. Since
the combination of two rotated zipper structures does not imply any
additional strain and fits seamlessly at the joints, we can easily
construct different facet lengths consisting of multiple zipper
units. This allows to compute the GB energies
\begin{equation}
  \label{eq:gb-energy}
  \gamma = \frac{E_\text{GB} - N e_\text{coh}}{A}
\end{equation}
for $T = 0$ and without externally-applied stress ($\sigma = 0$), with
$E_\text{GB}$ being the energy of a region consisting of $N$ atoms
containing the GB, and $e_\text{coh}$ being the cohesive energy per
atom of the bulk fcc phase (defined here as a negative number). The GB
area $A$ is the area projected onto the average GB plane, e.g., the
$(1\,10\,11)$ plane in Fig.~\ref{fig:motifs}(a), even for faceted
GBs. Figure~\ref{fig:facet-energies}(c) shows that the minimum GB
energy always occurs at the smallest facet length for our samples with
$\ang{46.83} \leq \theta \leq \ang{55.59}$, opposite to the effects
observed in many other faceted GBs \cite{Hamilton2003, Wu2009,
  Banadaki2016} and in our $\Sigma3$ GB [$\theta = \ang{60}$, dark
gray line in Fig.~\ref{fig:facet-energies}(c)]. In fact, energies
converge asymptotically to the value of zipper facets with zero
junction energy (infinite facet length), showing that nanofacets are
even stable compared to microscale facets. The same results are
obtained for Al and Ag (Supplemental
Fig.~\ref*{fig:suppl:facet-energies-symm-asymm}). As a further
verification step, we also conducted an MD simulation starting from
longer facets consisting of 11 zipper motifs and thus a facet length
of $L = \SI{9.9}{nm}$. In the periodic tilt axis direction we used a
thickness of \SI{6.3}{nm}. The simulation was performed at
\SI{500}{K}, to provide just enough thermal energy that the system can
evolve in the limited timescale of MD, while not probing the
high-temperature regime. The facet junctions of this system
immediately serve as nucleation sites for the appearance of nanofacets
with minimal facet length (Fig.~\ref{fig:md-shrinkage}). The GB is
fully nanofaceted after \SI{2.2}{ns}. Apart from confirming the
hypothesis that the domino phase can be regarded as a stable,
nanofaceted zipper variant, this also demonstrates that even a GB with
micrometer-sized facets would be unstable against facet length
reduction, because the facet refinement starts locally at the
junctions. It remains to explore why this phenomenon occurs in our
$[11\overline{1}]$ tilt GBs for $\theta < \ang{60}$.

\begin{figure*}
  \centering
  \includegraphics[]{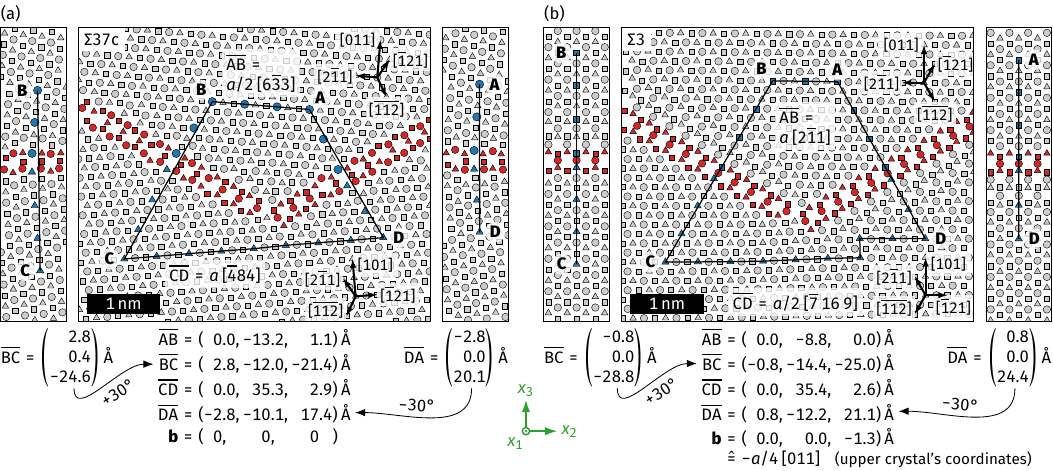}
  \caption{Burgers circuits around facet junctions in the (a)
    $\Sigma37$c and (b) $\Sigma3$ tilt GBs. Red atoms highlight the GB
    motifs. The symbols used for the atoms (circle, triangle, square)
    indicate that they belong to different $(11\overline{1})$
    planes. Blue atoms and black lines represent the closed Burgers
    circuit. The circuits are constructed as explained in the methods
    section: After indexing the crystallites, the lines
    $\overline{\mathrm{AB}}$ and $\overline{\mathrm{CD}}$ are
    constructed by following the interatomic distances
    $\Delta \mathbf{X}$ in the defect-free reference crystal along the
    indexed crystal directions; $\overline{\mathrm{AB}}$ in the
    coordinate system of the top crystal and $\overline{\mathrm{CD}}$
    in the coordinate system of the bottom crystal. They are then
    transformed into the common coordinate system $(x_1, x_2, x_3)$ as
    shown below the images in green. Values are listed in \AA{} below
    the images. The lines $\overline{\mathrm{BC}}$ and
    $\overline{\mathrm{DA}}$ are then marked in the system containing
    the facet junction and transposed into the junction-free systems
    containing only the left or right zipper structure (smaller images
    on the left and right). The line vectors are measured directly in
    these simulations from the atomic positions of points A, B, C, and
    D. The results are then rotated by $\pm \ang{30}$ around $x_1$. We
    then obtain $\mathbf{b} = \overline{\mathrm{ABCD}}$, which is (a)
    $\mathbf{b} = \mathbf{0}$ for the $\Sigma37$c GB and (b)
    $\mathbf{b} = a/4 [0\overline{1}\overline{1}]$ for the $\Sigma3$
    GB. As expected, the Burgers vector of the other facet junction in
    $\Sigma3$ has the opposite sign ($a/4 [011]$, analysis in
    Supplemental Fig.~\ref*{fig:suppl:Burgers-junction:others}).}
  \label{fig:burgers-circuits}
\end{figure*}

\subsection{Model}

In Eq.~5 of Ref.~\cite{Hamilton2003}, a criterion was derived for
energetic stabilization of finite facet lengths from interactions
between dislocation content and line forces at the junctions. The
facet junction energy consists of contributions from the junction's
core energy, dislocation interaction (due to the junction Burgers
vector $\mathbf{b}$), and line forces $f$ due to a discontinuity of
the GB excess stress $[\tau]$ at the junction. We can thus slightly
restate the energy of faceted GBs from Ref.~\cite{Hamilton2003} as
\begin{equation}
  \label{eq:faceted-gamma}
  \gamma = \frac{\gamma_\text{facet}}{\cos \phi}
           + \frac{E_\text{core}}{tL}
           + \left(k_1b^2 + k_2 f b - k_3 f^2 \right) \frac{\ln (L/\delta_0)}{L}
           .
\end{equation}
The first term is the GB energy of the facet's GB phase projected onto
the average GB plane, where $\phi$ is the inclination of the facet
(here $\phi = \ang{30}$, generally in the range
$[\ang{0};\ang{90}]$). The second term contains the facet junction's
core energy $E_\text{core}$, normalized by the facet junction line
length $t$ along the tilt axis and the facet length $L$ (distance of
junctions). The final term includes the dislocation--dislocation
interactions ($\propto b^2$), interactions between line forces
($\propto f^2$), and interactions between dislocations and line forces
($\propto f b$). The prefactors $k_i > 0$ contain elastic constants of
the material \cite{Eshelby1953, Foreman1955, Stroh1958,
  HirthLothe1992}. The length $\delta_0 \ll L$ is the core size of the
facet junction. A more detailed derivation (for GB phase junctions
instead of facet junctions, but mathematically equivalent) can be
found, e.g., in Ref.~\cite{Winter2022}, but is not required here.

Some clarification is, however, helpful to understand the implications
of this model. The two final terms are equivalent to the energy of a
low-angle GB (array of dislocations), except for the addition of the
line force $f$ \cite{Nabarro1952, Lejcek2010}. In the typical
derivation, the energy of a low-angle GB can be calculated under the
assumption that edge dislocations are arranged such that their
compressive and tensile strain fields overlap. This is the same as the
interaction of two parallel dislocations with opposite Burgers
vectors, which means that the strain energy per dislocation only has
to be calculated within a radius corresponding to the dislocation
distance, because it cancels out beyond that range \cite{Nabarro1952}.
In other words, dislocations introduce strain energy, but that strain
energy can be reduced (but not eliminated) by interaction with a
nearby dislocation with opposite Burgers vector
\cite{Nabarro1952}. The reduction is greater the closer the
dislocations are (more overlap of their opposite displacement
fields). Facet junctions also have to have alternating Burgers vectors
to preserve the total dislocation content of the GB. They can
therefore be treated the same way due to their alternating strain
fields \cite{Hamilton2003}. This means that while junction Burgers
vectors have an attractive force, each Burgers vector leads to a net
energy increase ($k_1b^2 \geq 0$). An increase of the facet length $L$
would therefore always lower the total number of facet junctions and
thus the total energy. GBs intrinsically also have a strain energy
cost due to their excess stress. If there is a discontinuity of this
excess stress at the facet junctions, the overlapping GB strain fields
partially compensate each other and the GB's strain energy is
reduced. This can be expressed as line forces that are opposite at
each junction, lowering the total energy ($k_3f^2 \leq 0$). More
detailed derivations have been provided for surface phase boundaries
that also exhibit such line forces \cite{Alerhand1988}. The
interaction $k_2fb$ does not play a role in our specific case, as
shown below.

Stabilization of nanofacets is thus a matter of competition between
these terms. Equation~\ref{eq:faceted-gamma} has an extremum at length
$L = \mathrm{e}\delta_0 \approx 0$ (due to $\delta_0 \ll L$). It is a
minimum if the prefactor of the final term is negative \footnote{It is
  $\frac{\mathrm{d}}{\mathrm{d}L} k'L^{-1}\ln(L/\delta_0) = k'L^{-2}
  (1-\ln{L/\delta_0})$. The function's single extremum is thus at
  $L = \mathrm{e}\delta_0$, where its second derivative
  $\frac{\mathrm{d}^2}{\mathrm{d}L^2} k'L^{-1}\ln(L/\delta_0) =
  k'L^{-3}(2\ln{L/\delta_0} - 3)$ is positive (condition for minimum)
  only for $k' = k_1b^2+k_2fb-k_3f^2 < 0$.}, i.e., if
\begin{equation}
  \label{eq:hamilton-criterion}
  f > b \frac{k_2 + \sqrt{k_2^2 + 4k_1k_3}}{2k_3}.
\end{equation}
Fulfilling this condition thus means that facet growth is predicted to
be disabled. A Burgers circuit around a junction in our faceted
$\Sigma37$c GB reveals that the Burgers vector content of the junction
with reference to the zipper phase is zero
[Fig.~\ref{fig:burgers-circuits}(a)]. This is consistent with the
combination of left and right zipper leading to an undistorted domino
phase motif [Fig.~\ref{fig:facet-energies}(a)] and is true for all of our
GBs with $\theta < \ang{60}$. Facet junctions in $\Sigma3$ GBs, in
contrast, contain a finite dislocation content
[Fig.~\ref{fig:facet-energies}(b) and
Fig.~\ref{fig:burgers-circuits}(b)]. To fulfill the criterion from
Ref.~\cite{Hamilton2003} for $b = 0$, we just need a positive $f$. In
the next step, we will therefore estimate $f$ for our GBs.

\begin{table}
  \caption{GB energy $\gamma$ and excess stress $[\tau_{ij}]$ for
    $\Sigma37$c and $\Sigma3$ GBs in Cu. The left and right zipper are
    degenerate states, as evidenced by their equal GB energy, but they
    can differ in the sign of their excess shear stress $[\tau_{12}$].
    Data for more misorientations, as well as for Al and Ag can be
    found in Supplemental Tables~\ref*{tab:suppl:excess-properties}
    and \ref*{tab:suppl:excess-properties-per-plane} and Supplemental
    Figs.~\ref*{fig:suppl:excess-properties}--\ref*{fig:suppl:excess-properties-per-plane}.}
  \label{tab:excess-stress}
  \begin{ruledtabular}
  \begin{tabular}{llD{.}{.}{1.3}D{.}{.}{2.2}D{.}{.}{2.2}D{.}{.}{2.2}}
    \multicolumn{2}{l}{structure}
        & \multicolumn{1}{c}{$\gamma$}
        & \multicolumn{1}{c}{$[\tau_{11}]$}
        & \multicolumn{1}{c}{$[\tau_{22}]$}
        & \multicolumn{1}{c}{$[\tau_{12}]$} \\
       && \multicolumn{1}{c}{(\si{J/m^2})}
        & \multicolumn{1}{c}{(\si{J/m^2})}
        & \multicolumn{1}{c}{(\si{J/m^2})}
        & \multicolumn{1}{c}{(\si{J/m^2})} \\
    \colrule
    $\Sigma37$c & domino       & 0.857 & -0.18 & +0.24 &  0.00 \\
                & left zipper  & 0.822 & +0.19 & -0.37 & +0.15 \\
                & right zipper & 0.822 & +0.19 & -0.37 & -0.15 \\
    \colrule
    $\Sigma3$   & left zipper  & 0.592 & -2.28 & -0.57 &  0.00 \\
                & right zipper & 0.592 & -2.28 & -0.57 &  0.00 \\
  \end{tabular}
  \end{ruledtabular}
\end{table}

We describe the GB excess stress as a tensorial value, defined at
$T = 0$ and $\sigma = 0$ as
\begin{equation}
    \label{eq:excess-stress}
    [\tau_{ij}] = \frac{\overline{\sigma}_{ij} V}{A},
\end{equation}
where $\overline{\sigma}_{ij}$ is the average residual stress in a
region of volume $V$ around the GB \cite{Frolov2012,
  Frolov2012a}. Indices $i,j = 1$ correspond to the tilt axis,
$i,j = 2$ to its orthogonal direction in the GB plane, and $i,j=3$ to
the GB normal. Excess stresses of $\Sigma37$c and $\Sigma3$ GBs are
listed in Table~\ref{tab:excess-stress}. Notably, the excess shear
stresses $[\tau_{12}]$ for the two zipper structures that make up the
domino structures have opposite values for all $\theta < \ang{60}$
(e.g., \SI{\pm0.15}{J/m^2} for $\Sigma37$c in Cu), while all other
components are the same. This can also be visualized by computing the
lattice strains (per-atom strain with reference to the perfect fcc
lattice) as calculated by the polyhedral template matching method
\cite{Larsen2016} in \textsc{ovito} \cite{Stukowski2010}: The shear
strains $E_{12}$ alternate between the facets (Supplemental
Fig.~\ref*{fig:suppl:S169-lattice-strains}).

The line forces can be computed as the sum of the two stress vectors
of the zipper facets. These vectors are calculated at an (imaginary)
surface separating the facets at a junction. This surface plane must
therefore be orthogonal to the average GB plane and have a normal
vector $\mathbf{v} = (0, 1, 0)$, representing the $x_2$ direction
along the GB (e.g., $[\overline{4}73]$ in the coordinate system of the
lower crystallite of the $\Sigma37$c domino GB) \cite{Mura1987,
  Winter2022}. Using Einstein notation we obtain
\begin{equation}
    \label{eq:line-force}
    f_j = \pm [\tau^\text{left}_{ij}] \cdot v_i \mp [\tau^\text{right}_{ij}] \cdot v_i.
\end{equation}
The two stress vectors at the separating plane point by definition in
opposite directions (analogous to action/reaction forces), leading to
opposite signs for the left and right junction terms. The overall sign of
the expression depends on the specific junction. Due to the opposite
sign of $[\tau_{12}]$ in both facets, this yields a nonzero value of
$f_1$ with alternating signs at each junction \footnote{It should be
  noted that the excess stresses in Eq.~\ref{eq:line-force} would in
  principle also have to be rotated by $\pm\phi$. However, the
  stresses are affected by the somewhat complex boundary conditions
  for which they are defined \cite{Frolov2012a}: The direction normal
  to the GB plane has open boundary conditions, while the other
  directions are fixed/periodic. The rotated stress tensor would thus
  have to be calculated with an inclined GB in a periodic box, which
  is not possible without introducing additional defects. We therefore
  settle here for a qualitative argument.}. The line force thus
contributes an energy proportional to $-f_1^2\ln (L/\delta_0)$ per
junction, where $L$ is the facet length \cite{Mura1987,
  Winter2022}. As stated above, this is a result of the reduction of
GB strain energy due to the alternating signs of the excess shear
stress. In other words, in our case there is no Burgers vector but
only opposite and therefore attractive line forces at the junctions
and the criterion for facet shrinkage is trivially fulfilled. This is
true for all of our GBs with misorientations $\theta <
\ang{60}$. Similar formalisms have been used for surface phase
coexistence \cite{Marchenko1981, Alerhand1988, Hannon2001,
  Zandvliet2004}. For the $\Sigma3$ GB, we find in contrast
$[\tau_{12}] = 0$ and $\mathbf{b} \neq \mathbf{0}$, leading to a
repulsion of the junctions and facet growth. The lines in
Fig.~\ref{fig:facet-energies}(c) are fits of
Eq.~\ref{eq:faceted-gamma} to the simulated GB energies, showing the
applicability of the model.

\subsection{Zipper structure and facet lengths}

\begin{figure}
    \centering
    \includegraphics{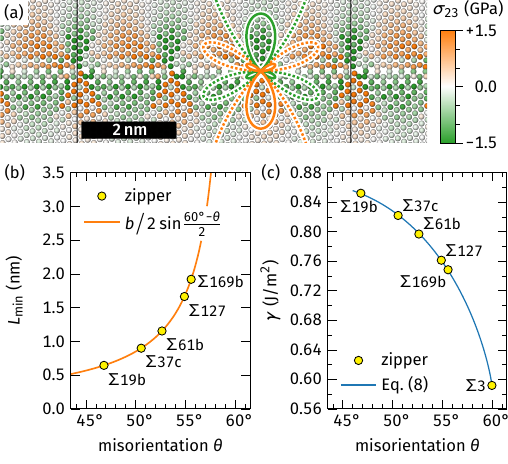}
    \caption{The trapezoidal unit as a virtual dislocation in the
      zipper structures with $\theta < \ang{60}$. If the trapezoidal
      unit has the character of an edge dislocation, the shear stress
      component $\sigma_{23}$ should be nonzero. The subfigure (a)
      shows the atomic stress of the zipper structure in the
      $\Sigma169$b tilt GB ($\theta = \ang{55.59}$), where the stress
      fields are most easily visualized.  We also plotted the contour
      lines of the analytical solution of the shear stress for an
      $a/6 \langle112\rangle$ edge dislocation
      \cite{HullBacon2011}. The dashed lines represent
      $\sigma_{23} = \pm \SI{0.25}{GPa}$ and the solid lines
      $\pm \SI{1}{GPa}$. The analytical solution fits to the observed
      stress (more stress components and an analysis of the $\Sigma3$
      tilt GB are shown Supplemental
      Fig.~\ref*{fig:suppl:trapezoidal-stress}). (b) Since the
      trapezoidal units have a Burgers vector (see also Supplemental
      Fig.~\ref*{fig:suppl:zipper-Burgers-circuits}), they compensate
      the difference in misorientation compared to the $\Sigma3$ GB
      ($\theta = \ang{60}$) and their distance can be calculated via a
      reversal of the Read--Shockley equation
      (Eq.~\ref{eq:read-shockley}). (c) The GB energy of the zipper
      can consequently be decomposed into the GB energy of $\Sigma3$
      plus the energy of a low-angle GB (Eq.~\ref{eq:zipper-energy},
      see Supplemental Fig.~\ref*{fig:suppl:zipper-energy} for
      additional data on Al and Ag).}
    \label{fig:zipper-as-mixed-GB}
\end{figure}

\begin{figure}
    \centering
    \includegraphics{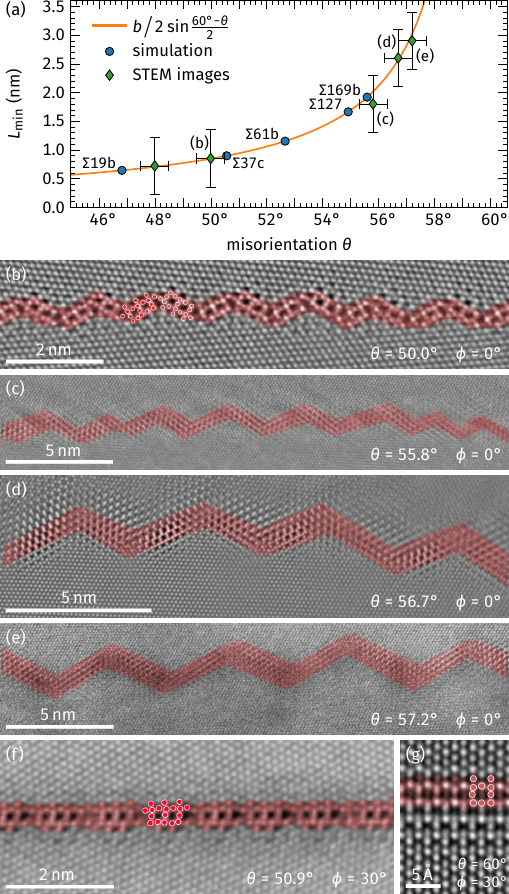}
    \caption{(a) Predicted facet lengths in domino assuming that the
      zippers are joined at the trapezoidal unit and that the
      trapezoidal unit is a virtual dislocation with
      $\mathbf{b} = a/6 \langle112\rangle$. Simulations represent the
      relevant domino phase, constructed by joining the left and right
      zipper structures. Experimental misorientations $\theta$ and
      facet lengths $L$ are measured from the STEM images in (b)--(e)
      and from Fig.~1 in Ref.~\cite{Meiners2020} for $\Sigma$19b. Note
      that the GB in (d) is likely not edge-on and therefore appears
      much wider than the other GBs. Data in (b) is adapted from
      Ref.~\cite{Langenohl2022} and data in (f) and (g) is adapted
      from Ref.~\cite{Langenohl2023} under the terms of the Creative
      Commons Attribution 4.0 International license. The measurement
      error of $\theta$ in all experimental images is \ang{\pm 0.5}.}
    \label{fig:facet-lengths}
\end{figure}

Regarding the zipper structure, the motifs in the $\Sigma3$ $\{112\}$
GB and close-by GBs with lower misorientation angles $\theta$ are
similar [Fig.~\ref{fig:facet-lengths}(f)--(g) and Supplemental
Figs.~\ref*{fig:suppl:S19b:zipper}, \ref*{fig:suppl:S37c:zipper},
\ref*{fig:suppl:S61b:zipper}, \ref*{fig:suppl:S127:zipper},
\ref*{fig:suppl:S169b:zipper}, \ref*{fig:suppl:S3:zipper}]: All GBs
contain the squares indicated in red. Additionally, GBs with
$\theta < \ang{60}$ contain another motif, indicated by the edge
dislocation symbol in Fig.~\ref{fig:motifs}(b)--(c). We call this the
trapezoidal unit \cite{Langenohl2023}. Inspection of the $\{112\}$
planes in Fig.~\ref{fig:motifs}(b) reveals that these motifs do indeed
resemble edge dislocations: It appears that $\{112\}$ planes terminate
at the trapezoidal unit. An analysis of the local stress fields
[Fig.~\ref{fig:zipper-as-mixed-GB}(a)] and a Burgers circuit analysis
[Fig.~\ref{fig:motifs}(c)] confirm that these units can be regarded as
virtual $a/6 \langle112\rangle$ dislocations when compared to the
reference state of the $\Sigma3$ GB~\cite{Langenohl2023}. It is
important to note that these are not real dislocations (in principle
the primary dislocation content required to obtain a misorientation
$\theta = \ang{50.57}$ for $\Sigma37$c is lower than to obtain
$\theta = \ang{60}$ for $\Sigma3$ \cite{Read1950}), but that they
behave similarly. We thus term them ``virtual dislocations''. The
energy of the zipper GB can therefore be written as a combination of
the $\Sigma$3 GB energy plus the energy of a low-angle GB as
\begin{equation}
  \label{eq:zipper-energy}
  \gamma(\theta) = \gamma_{\Sigma3} +
  \frac{Kb^2}{L_\text{min}(\theta)}
  \ln{\frac{L_\text{min}(\theta)}{\pi \delta_0}},
\end{equation}
where the second term is the energy of a low-angle GB
\cite{Nabarro1952, Lejcek2010}, defined by its dislocation interaction
energy \cite{Nabarro1952} with $K$ being an elastic constant
[$K = G/(4\pi - 4\pi\nu)$ for edge dislocations in isotropic materials with
shear modulus $G$ and Poisson's ratio $\nu$] and $\delta_0$ an
effective dislocation core size that also includes the core energy
term \cite{Han2018, SwethaArxiv}. The length $L_\text{min}$ represents
the distance between trapezoidal units.

This is relevant to the faceting because the distance between trapezoidal units can be written as
\begin{equation}
  \label{eq:read-shockley}
    L_\text{min}(\theta) = \frac{b}{2\sin\frac{\ang{60} - \theta}{2}}
\end{equation}
via a reversal of Read and Shockley's equation \cite{Read1950}, as
validated in Fig.~\ref{fig:zipper-as-mixed-GB}(b). The GB energy from
Eq.~\ref{eq:zipper-energy} with $K$ and $\delta_0$ as fit parameters
then also matches the computed GB energies
[Fig.~\ref{fig:zipper-as-mixed-GB}(c)]. Figure~\ref{fig:facet-energies}(a)--(b)
shows that the zipper facets join seamlessly at the trapezoidal unit,
and cannot join seamlessly if it is absent ($\Sigma3$). Thus,
$L_\text{min}$ is also the minimal facet length of the domino
structure, as confirmed in
Fig.~\ref{fig:facet-lengths}(a). Figures~\ref{fig:facet-lengths}(b)--(e)
show STEM images of domino phases with different misorientation angles
$\theta$. The simulations match the experimental observations where
comparison is possible. For larger $\theta$ values it becomes obvious
that the domino structure is faceted. The experimental facet lengths
correspond to the predictions [Fig.~\ref{fig:facet-lengths}(a)]. We
never observed longer facets in our data, despite annealing after
deposition for at least \SI{1}{h} at around
$T \approx \SI{700}{K} \sim 0.5 T_\text{melt}$. Faceting/defacting
transitions in $\Sigma3$ $\{011\}$ GBs in Al were previously observed
on the micrometer scale over holding times of \SI{40}{min} at similar
homologous temperatures \cite{Hsieh1989}, showing that the
facet-growth kinetics should definitely be sufficiently fast to
observe nanofacet growth if there were an energetic driving force. The
absence of long facets at atomic resolution in our experiment thus
provides proof for the stabilization of nanofacets.

\subsection{Asymmetric GBs}

\begin{figure}
    \centering
    \includegraphics{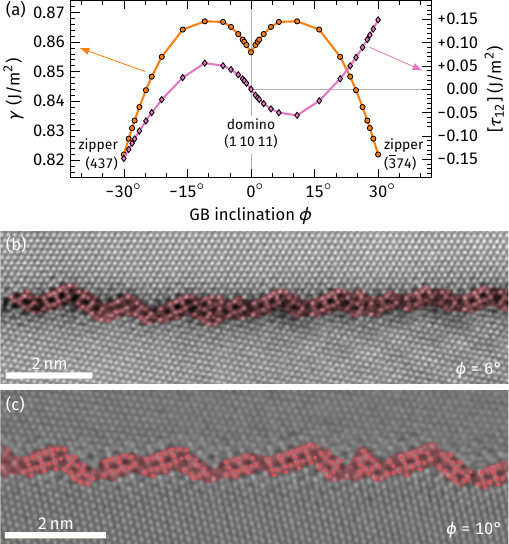}
    \caption{(a) Changing inclination for the $\Sigma37$c GB from the
      left zipper via domino to the right zipper by changing the
      length of one side of the facets. Inclination angle $\phi$ is
      relative to the domino plane. (b)--(c) STEM images of
      asymmetric $\Sigma37$c GBs. The measured inclination angle has an uncertainty of \ang{\pm0.5}.}
    \label{fig:asymmetric-energies}
\end{figure}

Combination of zipper structures into domino structures allows for a
simple method to produce asymmetric GBs in between the zipper and
domino inclinations by increasing the length of one of the facets. We
constructed different inclinations $\phi$ for $\Sigma37$c GBs and
calculated their GB energies and excess properties using molecular
statics. Figure~\ref{fig:asymmetric-energies}(a) shows that it is thus
possible to gradually transition from domino to zipper and vice versa
by changing the GB plane inclination. The combination of zipper
structures with opposite values of $[\tau_{12}]$ into domino with
$[\tau_{12}] = 0$ is also evident. We repeated these simulations for
Al and Ag, yielding equivalent results (Supplemental
Fig.~\ref*{fig:suppl:facet-energies-symm-asymm}). The energy landscape
suggests that domino and zipper can in principle transition into each
other.

\begin{figure*}
  \centering
  \includegraphics[]{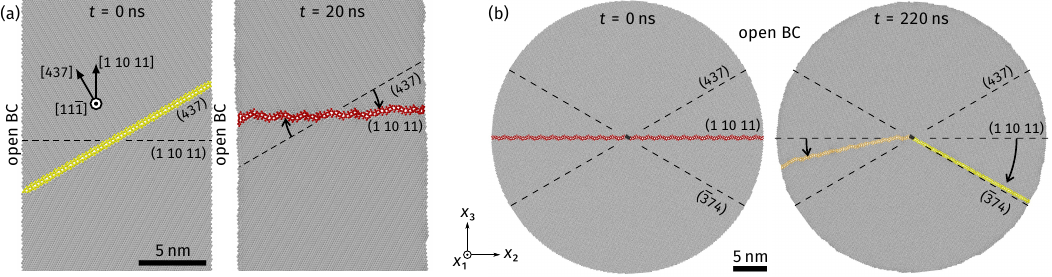}
  \caption{MD simulations of inclination changes driven by the GB
    energy. Annealing was performed at $T = \SI{600}{K}$, with
    subsequent cooling ($\dot{T} = \SI{e11}{K/s}$) to \SI{100}{K}. We
    used $\Sigma$37c GBs in Al, which do not show any transition to
    the pearl phase in this temperature range in contrast to Cu
    \cite{Brink2023}. Thickness along the periodic $x_1$ direction was
    \SI{1.4}{nm}. The small thickness was chosen to accelerate the
    kinetics of inclination changes. (a) When starting from a zipper
    GB (yellow) inclined by \ang{30} in a simulation cell with open
    boundary conditions in $x_2$ direction (left image), the GB can
    reduce its energy by rotating onto the (1\,10\,11) plane of the
    domino phase (red atoms, right image), despite the higher GB
    energy of domino compared to zipper. The reduction of GB area
    outweighs the normalized GB energy differences.  (b) We then
    performed simulations where the GB area does not change with the
    inclination by using a cylindrical setup. The spatial positions of
    the GB atoms in the center of the cylinder (black) were fixed to
    only allow GB rotation, but not movement normal to its plane. Now
    the GB is driven from the initial domino phase (red atoms, left
    image) to the zipper phase (yellow atoms, right image), purely due
    to the differences in GB energy. However, even after \SI{220}{ns}
    annealing time, the rotation on the left side is incomplete
    (orange atoms). This highlights that the small GB energy
    differences lead to small driving forces for rotation and
    generally relatively slow kinetics. These simulations demonstrate
    that inclination changes can be driven by GB energy but are highly
    dependent on the boundary conditions.}
  \label{fig:asymmetric-rotate}
\end{figure*}
For this, we performed two sets of MD simulations on $\Sigma$37c tilt
GBs in Al. We chose Al as an element because the pearl phase is not
stable below \SI{600}{K} \cite{Brink2023} and we can thus avoid a GB
phase transition into the pearl phase, which might suppress any
inclination change.  First, we build a simulation cell where the
horizontal plane (with normal $x_3$) corresponds to the GB plane of
the domino phase, shown in Fig.~\ref{fig:asymmetric-rotate}(a). We
inserted a zipper GB (which was consequently inclined by \ang{30}
towards the domino plane) and used open boundary conditions in the
$x_2$ direction, allowing the GB to rotate and thereby change its
length. While the GB energy of domino is higher than that of zipper
[Fig.~\ref{fig:asymmetric-energies}(a)], the rotation onto the
horizontal plane shortens the GB. Indeed, we observed that the system
reduced its total energy by transforming into the domino phase,
rotating by \ang{-30}. Second, we produced a cylindrical sample, such
that all inclinations lead to the same GB length and fixed the GB in
the center [Fig.~\ref{fig:asymmetric-rotate}(b)]. Thus, the GB can
rotate but cannot change its length or migrate. Here, GBs with
different inclinations slowly rotated towards the zipper
structure/inclination, which has the lowest GB energy per unit
area. This shows that inclination transitions are possible, but depend
strongly on the boundary conditions and are significantly more
complicated in realistic microstructures.

The fact that the nanofacets simultaneously act as pure steps on the
GB ($\mathbf{b} = \mathbf{0}$) and allow the construction of
asymmetric GBs raises the question of nomenclature. Indeed, GB facets
are typically much larger than the structures observed here, although
nanosized GB facets have been reported before \cite{Medlin2017,
  Peter2018}. Instead of using the concept of faceting, one could
therefore also consider the alternating facets as GB steps, especially
on the asymmetric GBs, or even call the domino structure a
reconstruction in analogy to surface science. We decided to call the
phenomenon faceting nonetheless because the domino motifs are clearly
two inclined zipper motifs and the construction of asymmetric GBs by
faceting is well established. It seems arbitrary to us to draw a line
between faceting and reconstruction on the basis of the facet length,
although the stability of our nanofacets might provide a distinction.

Finally, other types of structural defects could also potentially
compensate asymmetric inclinations and we do not claim that
lengthening facets is the only way to construct asymmetric
domino/zipper GBs. Nevertheless,
Figs.~\ref{fig:asymmetric-energies}(b)--(c) show STEM images of two
different inclinations whose atomic structures indeed consist of
longer and shorter facets, highlighting that low-energy configurations
do occur this way.

\section{Conclusion}

A set of $[11\overline{1}]$ tilt GBs in fcc metals exhibit nanofacets
that are energetically stable compared to longer facets at
misorientations $\theta < \ang{60}$. We call the atomic structure of
these GBs the domino structure. Atomic-resolution imaging is required
to resolve the nanofacets experimentally. The GBs are made of
zipper-type facets that can be joined seamlessly into domino
structures (Fig.~\ref{fig:motifs}). Their facet junction occurs at the
site of the zipper structure's trapezoidal motif---which behaves like
an $a/6 \langle 112 \rangle$ edge dislocation. The special feature of
this motif is that it allows facets to join without introducing an
additional facet junction Burgers vector. Only an attractive line
force due to the GB excess stress remains
(Eqs.~\ref{eq:faceted-gamma}--\ref{eq:line-force}). In agreement with
theory \cite{Hamilton2003}, this is sufficient to stabilize the
nanofacets energetically (Fig.~\ref{fig:facet-energies}). STEM
experiments on Cu thin films confirm the predicted, nanoscale facet
lengths for a range of misorientations
(Fig.~\ref{fig:facet-lengths}). Because the trapezoidal unit is not
present in the more well-studied $\Sigma3$ GBs, these GBs exhibit a
junction Burgers vector and thus a driving force for facet growth. The
faceting/defaceting in $\Sigma$3 \cite{Hsieh1989, Wu2009,
  Straumal2016} is thus likely a sign of a GB phase transition (change
away from the domino/zipper structures) and should be investigated in
the future.

\vspace{\baselineskip}

\begin{acknowledgments}
  We thank Gunther Richter and his team from the Max Planck Institute
  for Intelligent Systems for producing the Cu thin film by molecular
  beam epitaxy.
  This project has received funding from the European Research Council
  (ERC) under the European Union's Horizon 2020 research and
  innovation programme (Grant agreement No.~787446; GB-CORRELATE).

  T.B., G.D., and C.H.L. each conceptualized parts of the
  study. T.B. performed and analyzed the atomistic computer
  simulations, while S.P. analyzed the elastic interactions due to
  Burgers vectors. L.L. performed the experimental sample preparation,
  HAADF-STEM investigations, and analysis of the corresponding
  datasets. The project was supervised by C.H.L. and G.D., who also
  contributed to discussions. G.D. secured funding for T.B., L.L., and
  S.P. via the ERC grant GB-CORRELATE. T.B. and L.L. prepared the
  initial manuscript draft and all authors contributed to the
  preparation of the final manuscript.
\end{acknowledgments}

\end{document}